\documentclass[11pt]{article}
\usepackage{latexsym}
\usepackage{amssymb}
\usepackage{epsf}
\usepackage{amsmath}
\usepackage[hypertex]{hyperref}
\usepackage{graphicx}

\newcommand{\tr}{{\rm Tr}}

\begin{document}

\pagestyle{empty}

\begin{flushright}
LA-UR-07-7067\\
SLAC-PUB-12893
\end{flushright}

\vspace{2.5cm}

\begin{center}

{\bf\LARGE Sweet Spot Supersymmetry and Composite Messengers}
\\

\vspace*{1.5cm}
{\large 
Masahiro Ibe$^{1,2}$ and
Ryuichiro Kitano$^3$
} \\
\vspace*{0.5cm}

$^1${\it Stanford Linear Accelerator Center, Stanford University,
                Stanford, CA 94309} \\
$^2${\it Physics Department, Stanford University, Stanford, CA 94305}\\
$^3${\it Theoretical Division T-8, Los Alamos National Laboratory, Los Alamos, NM 87545}\\
\vspace*{0.5cm}

\end{center}

\vspace*{1.0cm}

\begin{abstract}
{\normalsize
\baselineskip 14pt

Sweet spot supersymmetry is a phenomenologically and cosmologically
perfect framework to realize a supersymmetric world at short
distance. We discuss a class of dynamical models of supersymmetry
breaking and its mediation whose low-energy effective description falls
into this framework.  Hadron fields in the dynamical models play a role
of the messengers of the supersymmetry breaking. As is always true in
the models of the sweet spot supersymmetry, the messenger scale is
predicted to be $10^5~{\rm GeV} \lesssim M_{\rm mess} \lesssim
10^{10}~{\rm GeV}$. Various values of the effective number of messenger
fields $N_{\rm mess}$ are possible depending on the choice of the gauge
group.

}
\end{abstract} 

\newpage
\baselineskip 18pt
\setcounter{page}{2}
\pagestyle{plain}
\section{Introduction}

\setcounter{footnote}{0}

If $N=1$ supersymmetry is hidden in nature, it helps us to understand
the hierarchy between the strength of gravity and weak interactions and
also the variety of matter fields and gauge forces in the standard
model.
It is, however, not straightforward to correctly hide supersymmetry at
low energy. Especially, there has been a trouble in making the Higgs
sector suitable for electroweak symmetry breaking, i.e., the
$\mu$-problem. Also, the smallness of flavor mixing and CP violation has
been considered as an unnatural aspect of the hypothesis.

Recently, the present authors carefully considered those problems,
including cosmological one, and found a simple and realistic framework
of supersymmetry breaking and mediation~\cite{Ibe:2007km}. A small
explicit breaking of the Peccei-Quinn (PQ) symmetry triggers
supersymmetry to break down, and it induces $\mu$-term through (a
generalized version of) the Giudice-Masiero
mechanism~\cite{Giudice:1988yz}. The explicit breaking term also makes
messenger fields massive via classical supergravity
effects~\cite{Kitano:2006wz}. Correct sizes of gaugino and sfermion
masses are obtained through their loop diagrams (gauge
mediation~\cite{Dine:1981za,Dine:1993yw,Dine:1994vc,Dine:1995ag}).
The dangerous proton-decay operators of the mass dimensions four and
five are forbidden by the PQ symmetry.
A mechanism of producing dark matter of the universe is built-in;
non-thermally produced gravitinos through the decay of the Polonyi field
naturally explains the correct abundance~\cite{Ibe:2006rc}. This
framework solves many known problems in supersymmetric model building,
and have no apparent phenomenological or cosmological shortcomings left.

The framework, {\it the sweet spot supersymmetry}, is written in the
language of the low-energy effective field theory. It is possible to
construct various explicit models within this framework as ultraviolet
(UV) completions, and each of those falls into a parameter point (or
region) in the sweet spot supersymmetry. The parametrization can be done
by four quantities: the number of messenger $N_{\rm mess}$, the
$\mu$-parameter, a gaugino mass, and the messenger scale $M_{\rm mess}$.
Once we specify those parameters, we can calculate the spectrum of
superparticles by a simple program described in
Ref.~\cite{Ibe:2007km}. Conversely, by measuring those quantities at low
energy experiments, we can obtain information on UV models.

In a recent paper~\cite{Nomura:2007cc}, an economical UV model in this
framework was proposed, that addresses the origin of the small explicit
breaking term of the PQ symmetry. (It is called U(1)$_H$ symmetry in
Ref.~\cite{Nomura:2007cc}.)  It is found that the term can be
non-perturbatively generated in a QCD-like theory (SU($N_c$) gauge
theory with $N_c$ flavors), and the quark fields in that supersymmetric
QCD play a role of the messenger fields. A parameter region of the sweet
spot supersymmetry is identified for this UV completion: $N_{\rm mess} =
5$ and $10^{11}~{\rm GeV} \lesssim M_{\rm mess} \lesssim 10^{13}~{\rm
GeV}$. The lower bound on the messenger scale is obtained from a
consistency of the analysis.

In this paper, we argue that this class of models predicts $M_{\rm mess}
\lesssim 10^{10}$~GeV, rather than $M_{\rm mess} \gtrsim 10^{11}$~GeV
where the analysis in Ref.~\cite{Nomura:2007cc} is meaningful.
Nevertheless, we find that there is a consistent effective description
in terms of hadron fields in that case. We can find a supersymmetry
breaking vacuum where gaugino/sfermion masses are generated by loop
diagrams of hadronic messenger fields instead of elementary quarks.
Generalizations with Sp($N_c$) and SO($N_c$) gauge theories are also
discussed.

\section{Model}

We briefly review the framework of the sweet spot supersymmetry, and
present a dynamical model of supersymmetry breaking which falls into
this framework.

\subsection{Sweet spot supersymmetry}

The Lagrangian of the sweet spot supersymmetry is written in terms of
the fields in the minimal supersymmetric standard model (MSSM) (chiral
superfields $\Phi_{\rm MSSM}$ and gauge fields $W_\alpha$), the
Goldstino field $S$, and the messenger fields $f$ and $\bar f$ that have
quantum numbers of the standard model gauge group.\footnote{The
Lagrangian of the sweet spot supersymmetry is presented in
Ref.~\cite{Ibe:2007km} as the one after integrating out the messenger
fields $f$ and $\bar f$. The original form contains terms in
Eqs.~(\ref{eq:gaugino}) and (\ref{eq:sfermion}) instead of those
involving $f$ and $\bar f$. They are, of course, equivalent.} It is
defined by the supergravity Lagrangian with a K\"ahler- and a
superpotential, $K$ and $W$:
\begin{eqnarray}
 K &=& \Phi_{\rm MSSM}^\dagger \Phi_{\rm MSSM}
+ S^\dagger S+ f^\dagger f + \bar f^\dagger \bar f
\nonumber \\
\label{eq:kahler}
&& - {(S^\dagger S)^2 \over \Lambda^2} 
+ \left(
{c_\mu S^\dagger H_u H_d \over \Lambda} + {\rm h.c.}
\right)
- { c_H S^\dagger S (H_u^\dagger H_u + H_d^\dagger H_d) \over
\Lambda^2}\ ,\\
 W &=& W_{\rm Yukawa} (\Phi_{\rm MSSM}) + m^2 S + k S f \bar f + w_0\ .
\label{eq:super}
\end{eqnarray}
In the K\"ahler potential, there are direct interaction terms between
the Goldstino fields $S$ and the Higgs fields $H_u$ and $H_d$ suppressed
by a `cut-off' scale $\Lambda$. ($c_\mu$ and $c_H$ are $O(1)$
coefficients.)\footnote{We assumed that at least first two generations
of the quark and lepton superfields are weakly coupled in a theory above
the scale $\Lambda$ in order to avoid flavor changing or CP violating
interactions suppressed by the scale $\Lambda$. For an explicit example
of such microscopic models, see Ref.~\cite{Ibe:2007km,Kitano:2006wm}.}
These interactions are responsible for generating the $\mu$-term and the
soft mass terms for the Higgs fields.
In the superpotential, a linear term of $S$ represents the source term
of the $F$-component of $S$. The interaction term between $S$ and
messenger fields $f$ and $\bar f$ is responsible for gauge mediation.
The constant term $w_0$ is needed to cancel the cosmological constant
such that $w_0 = m_{3/2} M_{\rm Pl}^2 = m^2 M_{\rm Pl}/ \sqrt 3$, where
$m_{3/2}$ is the gravitino mass.
This is the most general Lagrangian with the PQ-symmetry, $PQ(S)=2$,
$PQ(H_u) = PQ(H_d) = 1$, and $PQ(m^2)=-2$, where $m^2$ represents the
small explicit breaking parameter. Smallness of the supersymmetry
breaking scale and also of the $\mu$-parameter are controlled by this
parameter.

Obviously, there is a supersymmetric minimum in this model where
\begin{eqnarray}
 \langle S \rangle = 0\ ,\ \ \langle f \bar f \rangle = - m^2 / k\ .
\end{eqnarray}
However, we can find a local minimum with broken supersymmetry if the
value of $k$ is small enough~\cite{Kitano:2006wz}.
 From the above $K$ and $W$, we obtain a scalar potential for the $S$
field:
\begin{eqnarray}
 V(S) = m^4 \left(
{4 \over \Lambda^2} |S|^2
+ {k^2 N \over (4 \pi)^2} \log \left( {k^2 |S|^2 \over \Lambda^2 } \right)
\right)
- \left(
2 m_{3/2} m^2 S + {\rm h.c.}
\right)\ .
\label{eq:pote}
\end{eqnarray}
The logarithmic term is a loop correction from the interaction term, $k
S f \bar f$, and $N$ is a number of fields running in the loop. For
example, $N=5$ if the messenger fields $f$ and $\bar f$ transform as 5
and $\bar 5$ representations under SU(5) symmetry ($(3,1)_{-1/3} \oplus
(1,2)_{1/2}$ and $(\bar 3, 1)_{1/3} \oplus (1,2)_{-1/2}$ under the
standard model gauge group).
The linear term, $2 m_{3/2} m^2 S$, is a supergravity effect; this is a
soft supersymmetry breaking term associated with the linear term in the
superpotential in Eq.~(\ref{eq:super}).
Once we ignore the logarithmic term, the minimum is at
\begin{eqnarray}
 \langle S \rangle = {\sqrt 3 \Lambda^2 \over 6 M_{\rm Pl}}\ .
\label{eq:vev}
\end{eqnarray}
This makes the messenger fields massive, and thus stabilizes the $f \bar
f$ direction. Supersymmetry is broken by $F_S \simeq m^2 + k \langle f
\bar f \rangle= m^2$.
For a large value of $k$, however, this local minimum disappears because
the quantum correction becomes stronger than the supergravity effects.
As we will see in Section~\ref{sec:upper}, the condition that there is a
meta-stable supersymmetry breaking vacuum provides an upper bound on the
messenger scale, $M_{\rm mess} = k \langle S \rangle$.

By integrating out those massive messenger fields, we obtain terms
responsible for the gaugino and sfermion masses (gauge
mediation)~\cite{Giudice:1997ni}:
\begin{eqnarray}
 f_{\rm kin} \ni  - {N_{\rm mess} \over (4 \pi)^2 } 
\log S\  W^\alpha W_\alpha\ ,
\label{eq:gaugino}
\end{eqnarray}
for the gauge kinetic function, and
\begin{eqnarray}
 K \ni - {4 g^4 N_{\rm mess} \over (4 \pi)^4 } C_2(R) (\log |S|^2 )
  \Phi_{\rm MSSM}^\dagger \Phi_{\rm MSSM}\ ,
\label{eq:sfermion}
\end{eqnarray}
with $N_{\rm mess}$ the number of the messenger fields ($N_{\rm mess}=1$
for a pair of 5 and $\bar 5$ representations of SU(5)). 
With the non-vanishing value of $F_S$ and $\langle S \rangle$, we obtain
gaugino/sfermion masses through the above interaction terms.

There are two dimensionful parameters in this model: $\Lambda$ and
$m_{3/2} (=m^2/(\sqrt 3 M_{\rm Pl})= w_0 / M_{\rm Pl}^2)$. The
interesting discovery in Ref.~\cite{Ibe:2007km} is that there is a sweet
spot in the two dimensional parameter space ($m_{3/2},\Lambda$) where
everything works out fine.
The choice is $(m_{3/2},\Lambda)\sim(1~{\rm GeV},10^{16}~{\rm GeV})$
with which we obtain correct sizes of the $\mu$-term, gaugino and
sfermion masses, and the abundance of gravitino dark matter. The fact
that $\Lambda$ is at the grand unification scale is also an interesting
coincidence.

\subsection{A model of dynamical supersymmetry breaking}

A part of the above Lagrangian,
\begin{eqnarray}
 K &\ni& S^\dagger S - {(S^\dagger S)^2 \over \Lambda^2}\ , \nonumber \\
 W &\ni& m^2 S\ ,
\label{eq:susy-b}
\end{eqnarray}
provides an effective description of a quite general class of
supersymmetry breaking models.
As long as the Goldstino superfield (a field or a combination of fields
which gets $F$-component VEV) is weakly coupled in the actual
supersymmetry breaking model, the above Lagrangian is obtained by
integrating out other massive fields in the model.\footnote{In general,
if there is no (approximate) symmetry under which $S$ is charged in a
supersymmetry breaking model, there can be a cubic term in the K\"ahler
potential, $K \ni S^\dagger S^2 + {\rm h.c.}$. However, such terms can
be shifted away by an appropriate field redefinition $S \to S + c$. In
the case of the sweet spot supersymmetry, the presence of the
approximate PQ-symmetry is assumed. That restricts the form of K\"ahler
and superpotential to be the ones in Eqs.~(\ref{eq:kahler}) and
(\ref{eq:super}). } The $\Lambda$ parameter represents the strength of
the self interactions of the Goldstino and $m^2$ is the size of the
supersymmetry breaking.

We consider a simple example of such models where the superpotential
terms in Eq.~(\ref{eq:super}), $m^2 S + k S f \bar f$, are replaced with
a single term~\cite{Nomura:2007cc},
\begin{eqnarray}
 W \ni k S (F \bar F),
\label{eq:superFF}
\label{eq:mess}
\end{eqnarray}
where $F$ and $\bar F$ transforms as $5$ and $\bar 5$ representations of
an SU(5)$_F$ group (which contains the standard model gauge group as a
subgroup). They also have quantum numbers ($5$ and $\bar 5$) of another
gauge group SU(5)$_H$ which becomes strongly coupled at an energy scale
$\Lambda_{\rm dyn}$. In the above term, the indices of SU(5)$_H$ and
SU(5)$_F$ are contracted in $(F \bar F)$.

We assume here that the gauge interaction of SU(5)$_H$ becomes strong in
the regime where the masses of messenger particles, $M_{\rm mess}$, are
not important, i.e., $M_{\rm mess} < \Lambda_{\rm dyn}$. In fact, in
Ref.~\cite{Nomura:2007cc}, it has been claimed that this regime is
incompatible with the mechanism of supersymmetry breaking and mediation
since the meta-stable vacuum in Eq.~(\ref{eq:vev}) disappears. We show,
however, that we still have the gravitationally stabilized vacuum where
supersymmetry is spontaneously broken. Furthermore, we will see in the
next Section that $M_{\rm mess}$ is almost always lower than the
dynamical scale $\Lambda_{\rm dyn}$ for the vacuum to be meta-stable.

Below the scale $\Lambda_{\rm dyn}$, there is an effective description
of the theory in terms of meson $M_{ij} \sim F_i \bar F_j$ and baryon
fields $B \sim F^5$ and $\bar B \sim \bar F^5$. The indices $i,j ( = 1 -
5)$ are those of SU(5)$_F$.
The effective superpotential is given by
\begin{eqnarray}
 W = k S\cdot \tr M + X( \det M - B \bar B - (\Lambda_{\rm dyn}^2/5)^5)\ ,
\end{eqnarray}
where a Lagrange multiplier $X$ is introduced in order to ensure the
quantum modified constraint to be satisfied~\cite{Seiberg:1994bz}.
We can find a meta-stable vacuum in the meson branch, $\det M =
(\Lambda_{\rm dyn}^2/5)^{5}$. By solving the constraint around the point
$M_{ij} = \Lambda_{\rm dyn}^2 \delta_{ij} / 5$, we obtain
\begin{eqnarray}
 \tr M = \Lambda_{\rm dyn}^2 
+ {1 \over 2}{\tr \delta M^2 \over \Lambda_{\rm dyn}^2/5 } 
+ {B \bar B \over (\Lambda_{\rm dyn}^2/5)^4} + \cdots \ ,
\end{eqnarray}
where $\delta M$ is the traceless part of the matrix $M$. We have
neglected higher order terms in the field expansion.
The effective superpotential below the scale $\Lambda_{\rm dyn}$ is then
given by
\begin{eqnarray}
 W_{\rm eff} = k \Lambda_{\rm dyn}^2 S 
+ S
\left( {\hat k_M \over 2 }\tr \delta \hat M^2 + \hat k_B \hat B \hat{
 \bar B} 
\right)\ .
\end{eqnarray}
The fields $\delta \hat M$, $\hat B$, and $\hat{ \bar B}$ are
canonically normalized fields. Through this normalization procedure,
$O(1)$ uncertainties arise in the coupling constants $\hat k_M \sim \hat
k_B \sim k$.

The effective superpotential above is exactly the one in
Eq.~(\ref{eq:super}) by the identifications of $m^2 \sim k \Lambda_{\rm
dyn}^2$, $k \sim \hat k_M $, $f \sim \delta \hat M$ and $\bar f \sim
\delta \hat M$. The baryon fields do not contribute to the
gaugino/sfermion masses since they are singlet under the standard model
gauge group. The field $\delta\hat M$, on the other hand, transforms as
the adjoint representation under the SU(5)$_F$ flavor group. (The
quantum numbers under the standard model gauge group are $(8,1)_0$
$\oplus$ $(1,3)_0$ $\oplus$ $(3,2)_{-5/6}$ $\oplus$ $(\bar 3, 2)_{5/6}$
$\oplus$ $(1,1)_0$.) By integrating out those meson fields, the terms in
Eqs.~(\ref{eq:gaugino}) and (\ref{eq:sfermion}) are obtained with
$N_{\rm mess} = 5$. This is consistent with the anomaly matching
condition for U(1)$_{\rm PQ}$--SU(5)$_F$--SU(5)$_F$.

We need to make sure that the strong dynamics does not destabilize the
vacuum in Eq.~(\ref{eq:vev}). Through the interaction term, $k S F \bar
F$, it is expected to appear higher dimensional operators in the
K\"ahler potential such as
\begin{eqnarray}
 \delta K \sim {N \over (4 \pi)^2}{|k S|^4 \over \Lambda_{\rm dyn}^2}\ ,
\label{eq:eff-k}
\end{eqnarray}
where $N=25$.  However, the effect of this term is smaller than that of
the term $-(S^\dagger S)^2/ \Lambda^2$ in Eq.~(\ref{eq:susy-b}) if
\begin{eqnarray}
 k \lesssim 3 \times 10^{-3}
\left(
{N \over 25}
\right)^{-1/5}
\left(
{m_{3/2} \over 1~{\rm GeV}}
\right)^{1/5}
\left(
{\Lambda \over 1 \times 10^{16}~{\rm GeV}}
\right)^{-2/5}\ .
\label{eq:k-bound-2}
\end{eqnarray}
Here we have used a relation, $m^2 = k \Lambda_{\rm dyn}^2 = \sqrt 3
m_{3/2} M_{\rm Pl}$. As we will see later, the above condition is always
satisfied when $M_{\rm mess} < \Lambda_{\rm dyn}$ that we have already
assumed.

There is a lower bound on the messenger scale by a condition that the
messenger fields should not be tachyonic:
\begin{eqnarray}
 M_{\rm mess}^2 = \hat k_M^2 \langle S \rangle^2 > \hat k_M F_S\ .
\end{eqnarray}
Thus, we obtain
\begin{eqnarray}
 M_{\rm mess} = \hat k_M \langle S \rangle > {F_S \over \langle S
  \rangle}
= 3 \times 10^5~{\rm GeV} \left(
{m_{3/2} \over 1~{\rm GeV}}
\right)
\left(
{\Lambda \over 1 \times 10^{16}~{\rm GeV}}
\right)^{-2}\ .
\end{eqnarray}
We will examine in the next Section whether we have a consistent
parameter region.

\section{Upper bound on the messenger scale}
\label{sec:upper}

We derive an upper bound on the messenger scale from the stability of
the vacuum in Eq.~(\ref{eq:vev}). From the discussion, we will learn
that the messenger scale is almost always lower than the dynamical scale
$\Lambda_{\rm dyn}$, consistent with the assumption made in the previous
section.

In order to derive an upper bound on the messenger scale we first
consider a region with $M_{\rm mess} > \Lambda_{\rm dyn}$.
In this case, the quark fields $F$ and $\bar F$ can be integrated out
without considering the non-perturbative effects.
The phenomenon of supersymmetry breaking can be understood in a slightly
different way in this regime. Below the scale $M_{\rm mess}$, the theory
matches to the pure supersymmetric SU(5)$_H$ gauge theory. Eventually at
a scale $\Lambda_{\rm eff}$, the superpotential acquires a contribution
from the gaugino condensation, $W \ni \Lambda_{\rm eff}^3$. Now, by a
matching condition of the gauge coupling constant at the scale $M_{\rm
mess}$, we can see that this term has a dependence on the field value of
$S$: $\Lambda_{\rm eff}^3 = M_{\rm mess} \Lambda_{\rm dyn}^2 = k S
\Lambda_{\rm dyn}^2$. This is the linear term of $S$ in
Eq.~(\ref{eq:super}) which causes supersymmetry breaking by $F_S = m^2 =
k \Lambda_{\rm dyn}^2$~\cite{Nomura:2007cc}.

The quantum corrections to the scalar potential of $S$ can be calculated
perturbatively in the picture where $F$ and $\bar F$ are elementary
fields. It is simply the logarithmic term in Eq.~(\ref{eq:pote}) with
$N=25$.
A condition to have a local minimum in the scalar potential
(\ref{eq:pote}) is
\begin{eqnarray}
 {1 \over 3 M_{\rm Pl}^2} - {4 \over \Lambda^2 }
 {k^2 N \over (4 \pi)^2 }
 > 0\ ,
\end{eqnarray}
from which the bound on $k$ is obtained to be
\begin{eqnarray}
 k < 3 \times 10^{-3} 
\left( { N \over 25 } \right)^{-1/2}
\left( { \Lambda \over 1 \times 10^{16}~{\rm GeV} } \right)\ .
\label{eq:k-bound}
\end{eqnarray}
Therefore, with the VEV of $S$ in Eq.~(\ref{eq:vev}), we obtain the
upper bound on the messenger scale to be
\begin{eqnarray}
 M_{\rm mess} < 4 \times 10^{10}~{\rm GeV} 
\left( { N \over 25 } \right)^{-1/2}
\left(
{\Lambda \over 1 \times 10^{16}~{\rm GeV}}
\right)^3\ .
\label{eq:mess-bound}
\end{eqnarray}

On the other hand, the dynamical scale $\Lambda_{\rm dyn}$ has a
relation to the $m^2$ parameter:
\begin{eqnarray}
 m^2 = k \Lambda_{\rm dyn}^2 = \sqrt 3 m_{3/2} M_{\rm Pl}\ .
\end{eqnarray}
 From this, we obtain
\begin{eqnarray}
 \Lambda_{\rm dyn} = 4 \times 10^{10}~{\rm GeV}
\left(
{k \over 3 \times 10^{-3}}
\right)^{-1/2}
\left(
{m_{3/2} \over 1~{\rm GeV}}
\right)^{1/2}\ .
\label{eq:lam5}
\end{eqnarray}
 From Eqs.~(\ref{eq:k-bound}), (\ref{eq:mess-bound}) and
(\ref{eq:lam5}), we conclude that the messenger scale is lower than the
dynamical scale unless the bound in Eq.~(\ref{eq:k-bound}) is
saturated. Note that we cannot go far from the sweet spot values of
$m_{3/2}$ and $\Lambda$, otherwise the natural solution to the
$\mu$-problem is spoiled.

This discussion justifies the assumption $M_{\rm mess} \lesssim
\Lambda_{\rm dyn}$. In this regime, Eq.~(\ref{eq:k-bound}) should be
understood as a condition for the coupling constants $\hat k_M$ and
$\hat k_B$ rather than for the fundamental coupling constant $k$.
Then, by a relation $\hat k_M \sim \hat k_B \sim k$, the inequality in
Eq.~(\ref{eq:mess-bound}) just results in a consistency condition:
$M_{\rm mess} \lesssim \Lambda_{\rm dyn}$.
The bound in Eq.~(\ref{eq:k-bound}) (barring $O(1)$ ambiguities in the
relation between $k$ and $\hat k_M$) is identical to the previously
obtained constraint in Eq.~(\ref{eq:k-bound-2}) which ensures the
stability of the potential against corrections from the strong dynamics.

In summary, we have obtained a consistent region
\begin{eqnarray}
 10^5~{\rm GeV} \lesssim M_{\rm mess} \lesssim 10^{10}~{\rm GeV}\ ,
\end{eqnarray}
for the messenger scale, where the hadron picture is appropriate for the
analysis.
Note, however, that this prediction is generally true in any models of
the sweet spot supersymmetry.
The only non-trivial prediction of this model is $N_{\rm mess} = 5$. In
the next section, we examine the same class of models with different
strong gauge groups. We find those models predict different values of
$N_{\rm mess}$.

\section{Sp(${\mathbf N_{\mathbf c}}$) and SO(${\mathbf N_{\mathbf c}}$) models}

The mechanism of supersymmetry breaking and its mediation works also in
Sp($N_c$) and SO($N_c$) gauge theories instead of SU(5)$_H$. In order
for the matching condition, $\Lambda_{\rm eff}^3 = M_{\rm mess}
\Lambda_{\rm dyn}^2$, to hold, gauge groups are determined to be Sp(4)
or SO(12).

The discussion is almost the same for the Sp(4) case. We introduce $F$
and $\bar F$, that are $(8,5)$ and $(8,\bar 5)$ under Sp(4)$_H$ $\times$
SU(5)$_F$ group. Again, the standard model gauge group is a subgroup of
the SU(5)$_F$ global symmetry. We assume an interaction term, $W \ni k S
(F \bar F)$, where both Sp(4)$_H$ and SU(5)$_F$ indices are
appropriately contracted in $(F \bar F)$.
Below the dynamical scale of the Sp(4)$_H$ gauge theory, $\Lambda_{\rm
dyn}$, the theory is described by meson fields, $M$. The superpotential
is
\begin{eqnarray}
 W = k S \cdot \tr M_{F \bar F} 
+ X ( {\rm Pf} M - (\Lambda_{\rm dyn}^2/5)^5)\ ,
\end{eqnarray}
with $X$ a Lagrange multiplier~\cite{Intriligator:1995ne}. The meson
field $M$ is a $10 \times 10$ antisymmetric matrix:
\begin{eqnarray}
 M = \left(
\begin{array}{cc}
M_{FF} & M_{F \bar F} \\
-M_{F \bar F}^T & M_{\bar F \bar F}\\
\end{array}
\right).
\end{eqnarray}
The submatrix $M_{FF}$, $M_{\bar F \bar F}$, and $M_{F \bar F}$
transform as 10, $\bar{10} $, and $1+24$ under the SU(5)$_F$ flavor
group, respectively.
By solving the constraint and canonically normalizing the fields, we
obtain
\begin{eqnarray}
 W_{\rm eff} = k \Lambda_{\rm dyn}^2 S
+ S \left( \hat k_{FF} \tr (\hat M_{FF} \hat M_{\bar F \bar F}) 
+ {\hat k_{F \bar F} \over 2} \tr \delta \hat{M}_{F \bar F}^2 \right)\ .
\end{eqnarray}
The effective number of messengers are $N_{\rm mess} = 8$ in this case.

The case with an SO(12) gauge group is essentially the same, yet a
little bit more complicated. The quarks $F$ and $\bar F$ transform as
$(12,5)$ and $(12,\bar 5)$ this time. Below the dynamical scale, the
effective theory is a U(1) gauge theory with superpotential:
\begin{eqnarray}
 W = k S \cdot \tr M_{F \bar F}
+ (\det M - (\Lambda_{\rm dyn}^2/5)^{10} ) E^+ E^-\ ,
\end{eqnarray}
near a point $\det M = (\Lambda_{\rm dyn}^2/5)^{10}$. The fields $E^\pm$
are dyons~\cite{Intriligator:1995id}.
The meson field $M$ is a $10 \times 10$ symmetric matrix:
\begin{eqnarray}
 M = \left(
\begin{array}{cc}
 M_{FF} & M_{F \bar F} \\
 M_{F \bar F}^T & M_{\bar F \bar F} \\
\end{array}\right)\ .
\end{eqnarray}
The submatrix $M_{FF}$, $M_{\bar F \bar F}$, and $M_{F \bar F}$
transform as 15, $\bar{15} $, and $1+24$ under the SU(5)$_F$ flavor group,
respectively.
By turning on the VEV of $S$, the minimum of the potential is at 
\begin{eqnarray}
 \langle M_{F \bar F} \rangle = (\Lambda_{\rm dyn}^2/5) \delta_{ij}\ ,
\end{eqnarray}
\begin{eqnarray}
 \langle E^+ E^- \rangle = - {k S \over 2 (\Lambda_{\rm
  dyn}^2/5)^9}\ .
\end{eqnarray}
In this vacuum, the U(1) symmetry is Higgsed at the scale $(k \langle S
\rangle \Lambda_{\rm dyn})^{1/2}$ and the trace part of $M_{F \bar F}$
and $E^\pm$ obtain masses (or eaten by the U(1) gauge field) and
decouple. The effects of those massive particles on the $S$ potential
are always smaller than the term $K \ni -(S^\dagger S)^2/\Lambda^2$ for
$M_{\rm mess} < \Lambda_{\rm dyn}$.
Below the scale $(M_{\rm mess} \Lambda_{\rm dyn})^{1/2}$, the effective
superpotential is
\begin{eqnarray}
 W_{\rm eff} = k \Lambda_{\rm dyn}^2 S 
+ S \left( \hat k_{FF} \tr (\hat M_{FF} \hat M_{\bar F \bar F})
 + {\hat k_{F \bar F} \over 2 } \tr \delta \hat{M}_{F \bar F}^2
\right)\ .
\end{eqnarray}
Here, we canonically normalized fields.
Again, this is the superpotential of the sweet spot supersymmetry. As
anticipated, the effective number of messengers is $N_{\rm mess} = 12$
in this SO(12) model.

In fact, there is another branch in the SO(12) model where the
superpotential is given by
\begin{eqnarray}
 W = k S \cdot \tr M_{\rm F \bar F} + M_{ij} q_i^+ q_j^- \ ,
\end{eqnarray}
where $q$'s are monopoles. There is no supersymmetry breaking vacuum in
this branch. This is consistent with the fact that gaugino condensation
cancels and $W_{\rm eff} = 0$ in this case.

Although these models are similar to the IYIT model of supersymmetry
breaking~\cite{Izawa:1996pk}, there are essential differences. In the
IYIT model, we need to introduce gauge singlet fields for each flat
direction in order to kill all the supersymmetric vacuum.
In the model presented in this paper, we introduced only one singlet
field $S$. Therefore, there is a supersymmetric vacuum at $S=0$ since we
do not fix all the flat directions. However, by the help of an external
dynamics, i.e., supergravity interactions, $S$ can be stabilized away
from the supersymmetric vacuum.

\section*{Acknowledgments}

The work of MI was supported by the U.S. Department of Energy under
contract number DE-AC02-76SF00515.


\end{document}